\documentclass[pra,aps,amsfonts, amssymb, amsmath, showkeys, nofootinbib,twocolumn,longbibliography]{revtex4-1}
\usepackage[english]{babel}
\usepackage[utf8]{inputenc}
\usepackage[T1]{fontenc}
\usepackage[export]{adjustbox}
\usepackage[pdftex, pdftitle={Article}, pdfauthor={Author},colorlinks=true, allcolors=blue ]{hyperref} 

\usepackage[T1]{fontenc}
\usepackage{graphicx}
\usepackage{dcolumn}
\usepackage{bm}
\usepackage[utf8]{inputenc}
\usepackage{multirow}
\usepackage[T1]{fontenc}
\usepackage{graphicx}
\usepackage{dcolumn}
\usepackage{bm}
\usepackage{siunitx}
\usepackage{subcaption}
\usepackage[italicdiff]{physics}
\usepackage[utf8]{inputenc}
\usepackage[toc,page,title]{appendix}
\usepackage{graphicx}
\usepackage{subcaption}
\usepackage{mwe}
\usepackage[T1]{fontenc}
\usepackage{tikz}
\usepackage{graphicx}
\usepackage{adjustbox}
\usetikzlibrary{arrows, hobby}
\usepackage{wrapfig}
\usepackage{graphicx}
\usepackage{amsmath}
\usepackage{amssymb}
\usepackage{hyperref}
\usepackage{ mathrsfs }
\usepackage{indentfirst}
\usepackage{braket}
\usepackage{color}
\usepackage{empheq}
\usepackage{lipsum,adjustbox}
\usetikzlibrary{shapes}
\usetikzlibrary{plotmarks}
\usepackage{tikz}	
\usepackage[utf8]{inputenc}
\usepackage[T1]{fontenc}
\usepackage{mathptmx}
\usepackage{tikz}
\usepackage{graphicx}
\usepackage{adjustbox}
\usetikzlibrary{arrows, hobby}
\usepackage{wrapfig}
\usepackage{graphicx}
\usepackage{amsmath}
\usepackage{amssymb}
\usepackage{hyperref}
\usepackage{ mathrsfs }
\usepackage{indentfirst}
\usepackage{braket}
\usepackage{color}
\usepackage{empheq}
\usepackage{lipsum,adjustbox}
\usetikzlibrary{shapes}
\usetikzlibrary{plotmarks}
\usepackage{tikz}	
\usetikzlibrary{backgrounds,fit,decorations.pathreplacing}  

\tikzset{pics/.cd,
collector/.style={code={
\draw[fill=gray!20] (0,0.5) arc(90:-90:0.75cm and 0.5cm) -- cycle;}},
splitter/.style={code={\draw[ultra thick] (#1:{sqrt(1/2)}) --
(#1+180:{sqrt(1/2)});}},splitter/.default=135}

\tikzstyle{beamsplitter}=[fill=blue, fill opacity=0.2]

\makeatletter
\newcommand*{\rom}[1]{\expandafter\@slowromancap\romannumeral #1@}
\makeatother

\begin{document}
\title{Mitigation of Channel Tampering Attacks in Continuous-Variable Quantum Key Distribution}

\author{S. P. Kish$^1$}\email{sebastian.kish@data61.csiro.au}  \author{C. Thapa$^1$}  \author{M. Sayat$^2$}    \author{H. Suzuki$^1$}  \author{J. Pieprzyk$^1$}  \author{S. Camtepe$^1$} 

\affiliation{$^1$ Data61, CSIRO, Marsfield 2122, NSW, Australia.}
 \affiliation{$^2$ Department of Physics, The University of Auckland, Auckland 1010, New Zealand.}





\begin{abstract}

Despite significant advancements in continuous-variable quantum key distribution (CV-QKD), practical CV-QKD systems can be compromised by various attacks. Consequently, identifying new attack vectors and countermeasures for CV-QKD implementations is important for the continued robustness of CV-QKD. In particular, as CV-QKD relies on a public quantum channel, vulnerability to communication disruption persists from potential adversaries employing Denial-of-Service (DoS) attacks. Inspired by DoS attacks, this paper introduces a novel threat in CV-QKD called the Channel Amplification (CA) attack, wherein Eve manipulates the communication channel through amplification. We specifically model this attack in a CV-QKD optical fiber setup. To counter this threat, we propose a detection and mitigation strategy. Detection involves a machine learning (ML) model based on a decision tree classifier, classifying various channel tampering attacks, including CA and DoS attacks. For mitigation, Bob, post-selects quadrature data by classifying the attack type and frequency. Our ML model exhibits high accuracy in distinguishing and categorizing these attacks. The CA attack's impact on the secret key rate (SKR) is explored concerning Eve's location and the relative intensity noise of the local oscillator (LO). The proposed mitigation strategy improves the attacked SKR for CA attacks and, in some cases, for hybrid CA-DoS attacks. Our study marks a novel application of both ML classification and post-selection in this context. These findings are important for enhancing the robustness of CV-QKD systems against emerging threats on the channel.

\end{abstract}

\flushbottom
\maketitle

%

%
\thispagestyle{empty}

\section{Introduction}

Quantum key distribution (QKD) is a quantum technology that ensures secure key distribution between two parties.
Traditionally, the parties are called Alice (transmitter) and Bob (receiver).
The confidentiality of shared keys against a potential eavesdropper (Eve) is guaranteed by the laws of quantum mechanics. 
Once established, the keys are used to encrypt communication between the parties.
Photon-based QKD employs discrete variable (DV) encoding, using discrete properties like polarization, while coherent light-based QKD involves continuous variable (CV) encoding, utilizing the continuous attributes of the electromagnetic field, such as amplitude and phase, for secure key establishment \cite{Pirandola20}.
Thus, we have two classes of protocols: DV-QKD and CV-QKD.
The first DV-QKD protocol has been designed by Bennett and Brassard which is referred to as BB84 \cite{BENNETT20147}. It performs especially well for long distances \cite{bestqkd}. 
In contrast, CV-QKD protocols generate secret key bits at a high rate over short distances \cite{bestcvfibre, kish2023comparison}. 

Despite significant advancements in CV-QKD and its commercial availability, vulnerabilities persist due to imperfections in implementation. This leads to various side-channel attacks such as local oscillator (LO) intensity fluctuation, wavelength, calibration, and saturation attacks \cite{sidechannel, joug2013,xma, huang2013, huang2014, xma2013, qin2016, Zhao_2019, shao2022}. Consequently, identifying new attack vectors on implementations is necessary for the continued robustness of CV-QKD. Additionally, the public quantum channel used in QKD introduces susceptibility to disruption, with the denial-of-service (DoS) attack being a significant threat \cite{dosDV}. This attack, disrupting both classical and quantum channels, involves Eve introducing errors by sporadically blocking traffic in the quantum channel.  In particular, in the case of CV-QKD, Eve can modify the channel transmittance $T$ with a DoS attack to reduce the secret key rate (SKR) \cite{dosyaun,ldos}. Inspired by the DoS attack, we devise a novel threat, where Eve does not merely reduce but amplifies the transmittance $T$ which has implications for the SKR. This leads to the Channel Amplification (CA) attack on CV-QKD as a subset of channel tampering attacks which can be defined as any physical attack that modifies the channel transmittance. Detection and mitigation of these attacks is crucial for ensuring the robustness of CV-QKD systems and is the focus of this study.



Emphasizing the need for vigilance, there is a recent trend that applies machine learning (ML) techniques to detect side-channel attacks in CV-QKD. For example, there have been applications of ML to the detection of calibration attacks \cite{Mao_2020}, wavelength attacks \cite{He:20} and universal quantum attacks \cite{maomarkov,universalattack}. However, accurate ML models for detecting these attacks rely on extensive datasets with large feature vectors, often utilizing computationally intensive convolutional neural networks (CNNs). Our ML approach to channel tampering attack detection streamlines classification by employing a decision tree classifier and focusing on more relevant data features. 

Unlike other ML applications for attack detection in CV-QKD \cite{Mao_2020,He:20,maomarkov,universalattack}, we not only identify the accuracy of ML classification but also quantify the utility of a mitigation strategy centered on post-selection. This post-selection technique, also known as data clusterization, has been previously applied to free-space CV-QKD in \cite{Ruppert_2019, deq}, and here we find an application to mitigation of channel tampering attacks in a CV-QKD optical fibre implementation. To our knowledge of the current literature, our study marks a novel application of both ML classification and post-selection in this context.


In this work, we devise a physical channel tampering attack implemented in a CV-QKD optical fiber setup. We employ a parameter estimation and ML model for accurate attack classification with minimal resources. We also investigate mitigation strategies after classification. Our contributions entail:
\begin{itemize}
\item Introducing the CA attack in a CV-QKD implementation and providing identifiable channel tampering attacks. The knowledge of the existence of novel physical channel tampering attacks improves the security of CV-QKD. 
\item Implementing a supervised ML model that is lightweight and fast to classify diverse channel tampering attacks. The advantage of the ML model is high-accuracy classification.
\item Minimizing overhead by extracting features from LO intensity measurements. The advantage of this scheme is the reduced time for classification of channel tampering attacks and capability to determine frequency of attack $f_\text{attack}$.
\item Employing post-selection on classified attacks to enhance SKR and thereby mitigate the effects of the attack.
\end{itemize}

The work is organized as follows. In Section~\ref{sec_intro}, we introduce the CV-QKD protocol implementation in optical fibre and the parameter estimation procedure. Next, in Section~\ref{sec_quantum_security}, we introduce the quantum security analysis. 
In Section~\ref{sec_attacks}, inspired by the DoS attack, we detail the framework for channel tampering attacks and identify three possible types of attacks: CA, CA-DoS and DoS attacks. 
In Section~\ref{sec_ml}, we simulate real case scenarios for the three cases of attacks and develop an ML model for classification. We find a high accuracy, even with typical noise found in the implementation, but this drops when the noise is high. 
In Section~\ref{sec_postselection}, we employ post-selection of the classified channel tampering attacks and compare SKRs. We find that ML classification and post-selection improves the SKR. Next, in Section~\ref{sec_discussion}, we discuss our results, touching upon the feasibility of such an attack with current optical technologies and the limitations of our model. In the next section, we summarize and conclude our study.
\begin{figure*}[t!]
\centering
\includegraphics[scale=0.35]{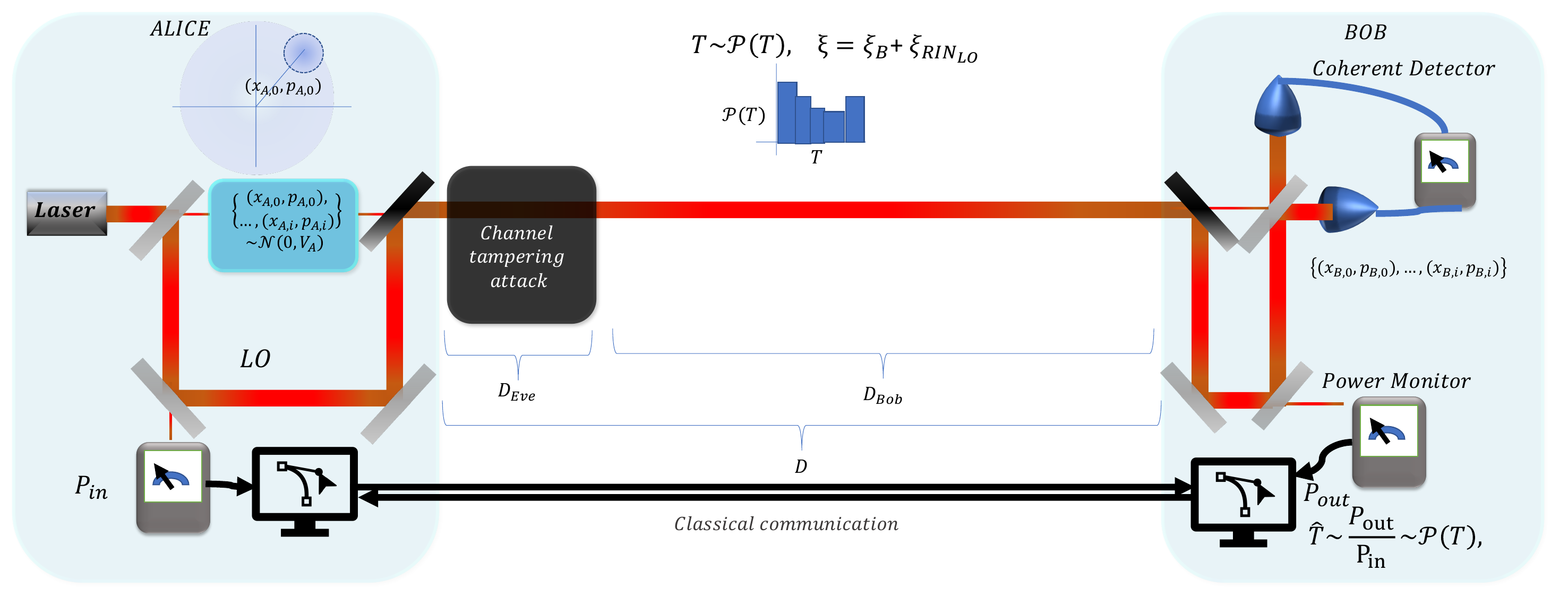}
\caption{An implementation of Gaussian modulated coherent state QKD protocol with channel tampering attacks in optical fibre. Alice splits a coherent laser source into a quantum signal and LO with an unbalanced beamsplitter. Alice modulates coherent signals with Gaussian distribution $\mathcal{N}(0,V_A)$. The signal and LO are multiplexed and sent through an optical fibre with transmittance $T$ and excess noise $\xi$. The channel tampering attack modifies the channel parameters up to a distance of $D_{\text{Eve}}$ and subsequently, leaves the rest of the optical fibre length of $D_{\text{Bob}}$ unchanged up to Bob. Bob measures the power fluctuations $P_{\text{out}}$ of the LO in addition to the coherent detector measurement. }
\label{qcomms}
\end{figure*}
\section{Parameter Estimation in Continuous Variable Quantum Key Distribution}\label{sec_intro}

We describe the Gaussian modulated coherent state QKD protocol \cite{gross1,gross2,cvhet}. In this protocol, Alice prepares coherent states $\ket{\alpha_{A_0} \ldots,\alpha_{A_i}}$ in the phase space described by the position quadrature $\hat{X}=a+a^\dagger$ and momentum quadrature $\hat{P}=-i(a-a^\dagger)$ where $a$ and $a^\dagger$ are the annihilation and creation operator, respectively. Alice modulates the coherent states according to a Gaussian distribution $\mathcal{N}((X_A=0,P_A=0),V_A N_0)$ centered at $X_A=0,P_A=0$ with variance $V_A N_0$ where $N_0$ is the shot noise, $V_A$ is Alice's modulation variance, $X_A$ is the mean of $\hat{X}$ and $P_A$ is the mean of $\hat{P}$. The coherent states with phase space positions at $\{ \{x_{A,0}, p_{A,0}\}, \ldots \{x_{A,i}, p_{A,i}\} \}$ are sent through the channel with transmittance $T$ and excess noise $\xi$. Bob receives the coherent states and measures the quadratures $\hat{X}$ and $\hat{P}$ using a heterodyne detector. The states received by Bob are correlated with the states sent by Alice. Bob can perform statistics on the received measurement data $\{ \{x_{B,0}, p_{B,0} \}, \ldots \{x_{B,i}, p_{B,i}\} \}$, which are also distributed according to a Gaussian distribution $\mathcal{N}((X_B=0,P_B=0),V_B)$.

In a standard CV-QKD implementation, parameter estimation is performed using a fraction of the quadrature measurements and the maximum likelihood estimation of $T$ and $\xi$. In our CV-QKD system depicted in Fig. \ref{qcomms}, Alice and Bob also continuously monitor the intensity of the LO and estimate $\hat{T}$.

In our CV-QKD system, we assume that the local oscillator (LO) locking the quadrature phases is polarization and/or time-multiplexed with the quantum signal. The design is similar to  \cite{huangsci} but with a few additions. Namely, a power monitor is placed at Alice and Bob to measure the optical power into the channel $P_{\text{in}}$ and out of the channel $P_{\text{out}}$. $P_{\text{in}}$ experiences power fluctuations due to temperature fluctuations and electronics. For Bob to estimate $\hat{T}$, Alice needs to constantly send $P_{\text{in}}$ using the classical communication channel. Alternatively, as this could be a security risk, Alice could characterize the mean and variance of $P_\text{in}$ before the protocol starts.
The LO experiences relative intensity noise $RIN$ due to the channel, which we quantify by  $\sigma_{\text{RIN,LO}}$ given by noise models in \cite{laudenbach}. 
The excess noise of the system is $\xi=\xi_B+\xi_{\text{RIN,LO}}$, where $\xi_{\text{RIN,LO}}=\frac{V_{\perp RIN,LO}}{4} \sigma_{\text{RIN,LO}}$ where we have defined $\xi_B$ to be the baseline excess noise without the RIN of the LO and $V_{\perp RIN,LO}=V_A+1$ is the variance of the quadrature without RIN of the LO.

Alice and Bob then perform parameter estimation, post-processing, and privacy amplification to obtain a positive SKR (see Ref. \cite{zhang2024continuousvariable} for a review on CV-QKD). In the usual parameter estimation, the measurement data at Bob is modeled by a normal linear transmission model:

\begin{equation}
    x_B=t x_A+z_R,
\end{equation} where for constant transmittance $t=\sqrt{T \eta}$, $\eta$ is the detector efficiency, $x_A$ are coherent states modulated by Alice and $z_R$ is the Gaussian noise of $\mathcal{N}(0,\sigma^2_R)$ where $\sigma^2_R=v_{\text{el}}+N_0+\eta T \xi$ is the channel noise. However, if $T$ is not constant i.e. $\hat{T}=\hat{t}^2/\eta$, as would be the case in a fluctuating channel, the expectation values $\mathbb{E}(T)$ and $\mathbb{E}(\sqrt{T})$ need to be estimated. A fraction $m/N$ of the total quadrature measurement data $N$ is used to estimate the parameters:

\begin{equation}
    \hat{t}=\sqrt{\eta} \mathbb{E}(\sqrt{T})=\frac{\frac{1}{m}\sum^{m-1}_{i=0} x_{A,i} x_{B,i}}{\frac{1}{m}\sum^{m-1}_{i=0} x^2_{A,i}},
    \label{sqrtE}
\end{equation}

and 

\begin{equation}
    \hat{\sigma}^2_R=\frac{1}{m} \sum^n_{i=0} (x_{B,i}-\hat{t} x_{A,i})^2.
\end{equation}



For the estimation of $T$, we note that the variance of Bob's measured quadrature $X_B$ is
\begin{equation}
    \text{Var}(X_B)=\mathbb{E}(X_B^2)=\eta \mathbb{E}(T) (V_A +\xi)+N_0+v_{\text{el}},
\end{equation} where $v_{\text{el}}$ is the electronic noise of Bob's detector. The cross-correlations or covariance between Alice and Bob are given by:

\begin{equation}
    \text{Cov}(X_A,X_B)=\mathbb{E}(X_A X_B)=\frac{1}{m}\sum^{m-1}_{i=0} x_{A,i} x_{B,i}=\mathbb{E}(\sqrt{\eta T}) V_A,
\end{equation}

Based on the measurement of the shot noise $N_0$, electronic noise $v_{\text{el}}$, Alice's modulation variance $V_A$ and excess noise $\xi$, one could estimate the mean value of $T$ by 
\begin{equation}
    \mathbb{E}(T)=\frac{\mathbb{E}(X^2_B)-N_0-v_{el}}{\eta V_A N_0+\eta \xi}.
\end{equation}
In Fig. \ref{qcomms}, Eve performs the channel tampering attack. In this attack, she modifies the transmittance parameter and its statistics with probability distribution $\mathcal{P}(T)$. Eve hijacks part of the optical fibre communication up to a length of $D_{\text{Eve}}$ and leaves the rest of the fibre length of $D_{\text{Bob}}$ unchanged. The parameters under attack of the modified channel are \cite{li2018} (which are also derived in the Appendix \ref{modified}):
\begin{equation}
    \begin{split}
        \hat{T}&=[\mathbb{E}(\sqrt{T})]^2, \\
        \hat{\xi}&=\frac{\mathbb{E}(T)}{[\mathbb{E}(\sqrt{T})]^2} (V_A+\xi)-V_A.
    \end{split}
    \label{parameterattack}
\end{equation}

The above approach relies on measuring other parameters precisely. On the other hand, monitoring the LO optical power provides more information about the distribution of $\sqrt{T}$ and $T$ without sacrificing an additional part of the key to monitor for channel tampering attacks. Monitoring the LO intensity has previously been suggested as a defense against wavelength and calibration attacks \cite{huang2013,joug2013}. In our CV-QKD protocol depicted in Fig. \ref{qcomms}, we assume that monitoring $T$ via the LO intensity is correlated with the estimated parameters in Eq. (\ref{parameterattack}).   


As Bob estimates $\hat{T}$ from the measurement of the LO intensity, a fraction of the quadrature measurement data $m/N$ is needed for parameter estimation of the normal unattacked $T$ and the excess noise $\xi$. For increased security, one could also determine correlations between $\hat{T}_{\text{LO}}$ from the LO and $\hat{T}_{\text{sig}}$ from quadrature measurements. In practice, there are statistical effects due to the finite number of symbols sent between Alice and Bob. As such, there is a worst-case scenario estimation for the channel parameters given by \cite{lev}:
\begin{equation}
    \begin{split}
        T_{wc} & \simeq T - w \sigma_T, \\
        \xi_{wc} & \simeq \frac{T}{T_{wc}}\xi+w \sigma_\xi, 
    \end{split}
    \label{finiteparam}
\end{equation} where

\begin{equation}
    \begin{split}
        \sigma_T &=\frac{2 T}{\sqrt{V_0 m}} \sqrt{c_{PE}+\frac{\xi+\frac{V_0+v_{\text{el}}}{\eta T}}{V_A}}, \\
        \sigma_\xi &= \sqrt{\frac{2}{V_0 m}}\frac{\eta T+V_0+v_{el}}{\eta T_{wc}}, 
    \end{split}
\end{equation} where $V_0=2$ for heterodyne detection, $m$ is the number of keys for the parameter estimation, $c_{PE}=0$ as in \cite{lev} and $w=\sqrt{2} \text{erf}^{-1}(1-\varepsilon_{PE})$ where $\varepsilon_{PE}$ is the error probability of parameter estimation failure.

\section{Quantum Security Analysis}\label{sec_quantum_security}
Eve's information is obtained by the Holevo bound $\chi_{\text{EB}}$,
which is determined by the covariance matrix of Alice and Bob. The SKR of CV-QKD in the asymptotic limit is \cite{laudenbach}
\begin{equation}
    K_{\infty}=\beta I_{\text{AB}}-\chi_{\text{EB}},
    \label{skr}
\end{equation} where $\beta$ is the reconciliation efficiency, and $I_{AB}$ is the mutual information between Alice and Bob. The SKR calculations in the asymptotic limit of the Gaussian modulated coherent state with heterodyne detection are shown in Appendix~\ref{appendixgg02}. Security proofs in the asymptotic limit are based on the reduction of coherent attacks to collective attacks and Gaussian optimality \cite{wolf,patron,nava}.
With finite-size effects, the equation for the secret key against collective (Gaussian) attacks for success probability of error correction $p_{\text{ec}}$ and $\varepsilon$-secrecy is given by \cite{limpirandola,charfinite}:
\begin{equation}
    K \ge \frac{n p_{\text{ec}}}{N} (K_\infty(T_{wc},\xi_{wc})-\frac{\Delta_{\text{aep}}}{\sqrt{n}}+\frac{\Theta}{n}),
\end{equation} for total block-size of $N$ and remaining symbols for the secret key $n=N-m$, where 

\begin{equation}
    \begin{split}
        \Delta_{\text{aep}}&:=4 \log_2{(2 \sqrt{d}+1)} \sqrt{\log_2\frac{18}{p^2_\text{ec} \varepsilon_s^4}}, \\
        \Theta &:=\log_2[p_{\text{ec}}(1-\varepsilon_s^2/3)]+2 \log_2\sqrt{2} \varepsilon_h,
    \end{split}
\end{equation} where $d$ is the size of the alphabet after analogue-to-digital conversion (usually set to $d=2^5$) and secrecy $\varepsilon=2p_{\text{ec}} \varepsilon_{\text{PE}}+\varepsilon_{\text{cor}}+\varepsilon_h+\varepsilon_s$.

\section{Channel Tampering Attacks}\label{sec_attacks}
As in Fig.~\ref{qcomms}, we split the Eve's capability into channel tampering attacks and quantum attacks. Eve has the capability of modifying $T$ of the channel. In this particular setup,  we assume that the signal and LO are equally affected by the modified channel. The latter is a fair assumption if the signal and LO are multiplexed in the same physical fibre. Subsequently, Eve has the capability of performing a quantum attack based on the modified channel parameters. { We emphasize that the channel tampering attack affects the parameter estimation of a complete CV-QKD protocol, which is different from an attack on the implementation of the quantum processes.} 

Normally, DoS attacks manipulate $T$ according to either a two-point or uniform distribution \cite{li2018}. In this attack implementation, we will only refer to the two-point distribution. The two-point distribution corresponds to intermittently blocking communications, with the transmittance, which varies from $0$ to $T_0$ and follows a binomial distribution of $B(1,p)$, where $p$ is a parameter between $0$ and $1$ chosen by Eve. 
The expected values are  $\mathbb{E}(\sqrt{T})=\mathbb{E}(B(1,p))\sqrt{T_0}=p \sqrt{T_0}$ and $\mathbb{E}(T)=\mathbb{E}(B(1,p)) T_0=p T_0$ where the expectation value of the binomial distribution is $\mathbb{E}(B(1,p))=p$.

Let Eve split the fibre with a loss $L$ (in dB/km) and replace it with a fibre with a lower loss $L'$. For example, Eve can patch the optical fibre during a blackout with an optical switch into a lower-loss fibre. Therefore the transmittance is $g T_0$, where $g=10^{-L' d_{\text{Eve}}/10}/10^{-L d_{\text{Eve}}/10}=10^{\Delta L d_{\text{Eve}}/10}$, $d_{\text{Eve}}$ is Eve's fibre cable length and $\Delta L=L-L'$. 
Thus, the above expected values become: 

\begin{equation}
\begin{split}
    \mathbb{E}(\sqrt{T})&=p \sqrt{g T_0}, \\
    \mathbb{E}(T)&=p T_0 g.
\end{split}
\label{twopoint}
\end{equation}
We can simulate such an attack by generating $T$ according to the distribution:
\begin{equation}
    \mathcal{P}(T) =g T_0 B(1,p) \mathcal{N} (T;T_{\text{Bob}}, \sigma_{RIN},0,1),
\end{equation} where $T_{\text{Bob}}=10^{-L d_{\text{Bob}}/10}$ with $d_{\text{Bob}}$ being the length of fibre from Eve to Bob and $\mathcal{N} (T;T_{\text{Bob}}, \sigma_{RIN},0,1)$ is a normal distribution between $0$ and $1$. In this model, Eve has the freedom to launch the CA attack at any point of the optical fibre. 
Note that a swap of optical fibres is not the only method to perform this attack by Eve. 
She can also change the vacuum properties of the channel. 
This can be done, for instance, by decoupling the signal and LO from the fibre into a vacuum chamber of refractive index $n_0=1$ and back into the fibre \cite{Mironov_2020}. 
\begin{figure*}[t!]
\centering
\includegraphics[width=0.8 \textwidth,trim=0mm 0mm 0mm 0mm, clip=true]{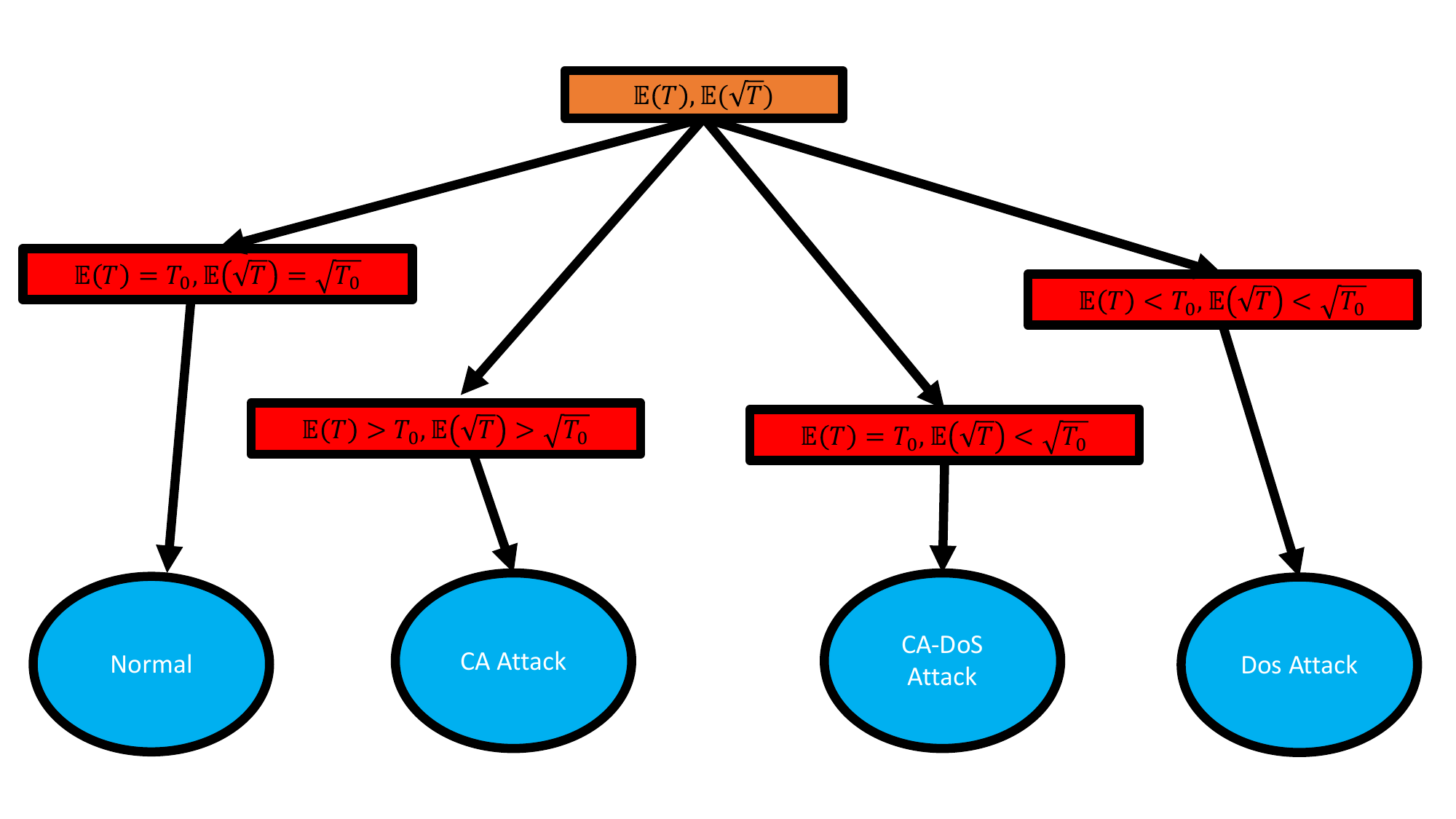}
\hfill
\caption{Decision tree classifier for classification of channel tampering attacks. The supervised ML model
learns from the labelled data the type of
channel tampering attack that occurred. It makes a decision based on the features of the data. }
\label{decisiontree}
\end{figure*}
Given the channel tempering attack, the estimated channel parameters become:
\begin{equation}
    \begin{split}
        T'&=p^2 g T_0 \\
        \xi'&=\frac{1}{p} (V_A+\xi)-V_A
    \end{split}
    \label{CTattack}
\end{equation}
%
Based on these parameters, we can simulate a realistic attack with a loss $L=0.2 ~\text{ dB/km}$ for standard fibre and $L'=0.15~\text{ dB/km}$ for ultra-low loss fibre. 
From this point forward, we define the CA attack with these parameters.

We consider the following case studies for Eve's channel tampering attacks:

{\bf Case \rom{1}}: $p=1$ and $g>1$. A successful CA attack is when the transmittance parameter $\hat{T}=g T_0$ is amplified, and excess noise remains the same $\hat{\xi}=\xi$. 

{\bf Case \rom{2}}: $p=1/\sqrt{g}$ and $g>1$. In this hybrid CA-DoS attack, the two-point distribution parameter is tuned by Eve such that $\hat{T}=T_0$ and $\hat{\xi}=\sqrt{g} (V_A+\xi)-V_A$. The excess noise is higher than if there were no attack but the average transmittance remains the same. 

{\bf Case \rom{3}}: $p<1$ and $g \le 1$. In this DoS attack, Eve performs a stronger DoS attack that reduces the SKR significantly, and therefore it reduces the number of key bits exchanged between Alice and Bob. 

Without the knowledge that the first type of attack in Case \rom{1} had occurred, a naive defense strategy would be to average over the channel parameters and follow information-theoretic security in Section \ref{sec_quantum_security}. In Case \rom{2}, if the attack is weak $p \approx 1$ and $g \approx 1$, Alice and Bob can continue
their interactions as usual, regardless of whether the attack is detected or not. 
Alternatively, if Eve is nearly blocking all communications as in Case \rom{2} \& \rom{3} i.e. when $p<<1$ and upon the classification of this attack, Alice and Bob can use a part of their key to randomly select a different optical fibre channel if available. 

We shall see in the next sections that the effect of the channel tampering attacks in Case \rom{1} \& \rom{2} on the SKR can be detected and mitigated.

\begin{figure*}[t!]
\centering
\begin{subfigure}{0.47\textwidth}
    \includegraphics[width=\textwidth,trim=0mm 50mm 20mm 50mm, clip=true]{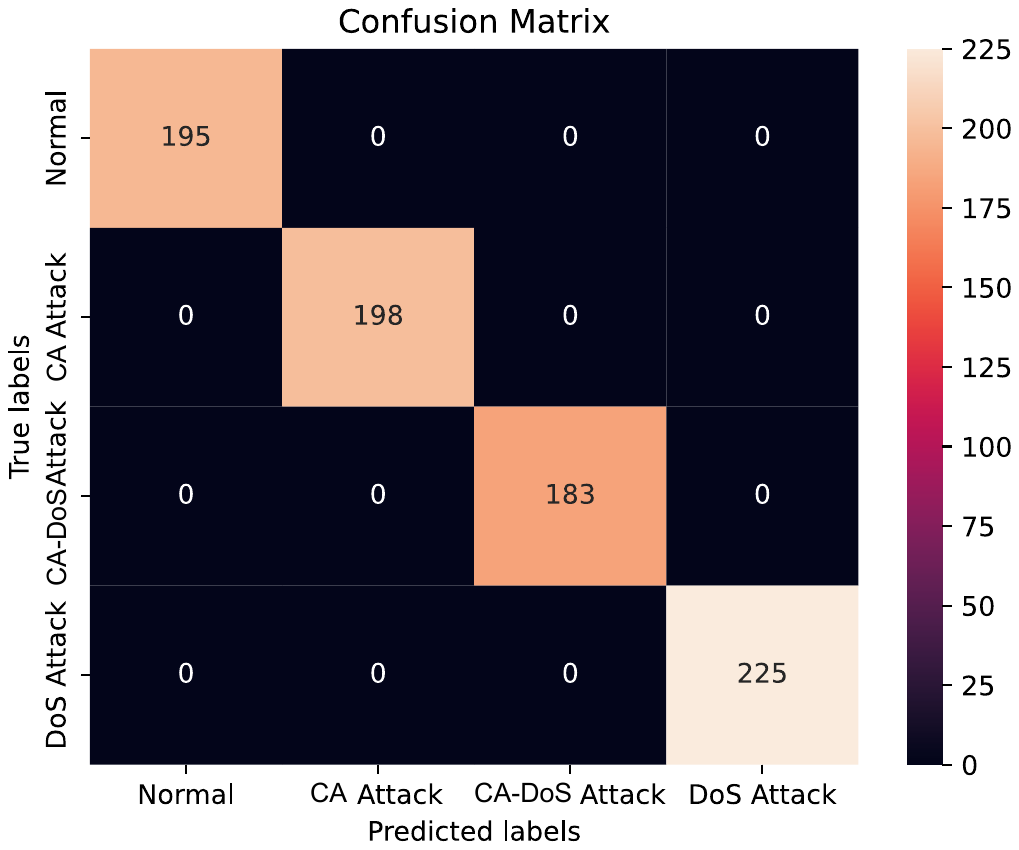}
    \caption{Accuracy= $100\%$.}
    \label{fig:first}
\end{subfigure}
\hfill
\begin{subfigure}{0.47\textwidth}
    \includegraphics[width=\textwidth,trim=0mm 50mm 20mm 50mm, clip=true]{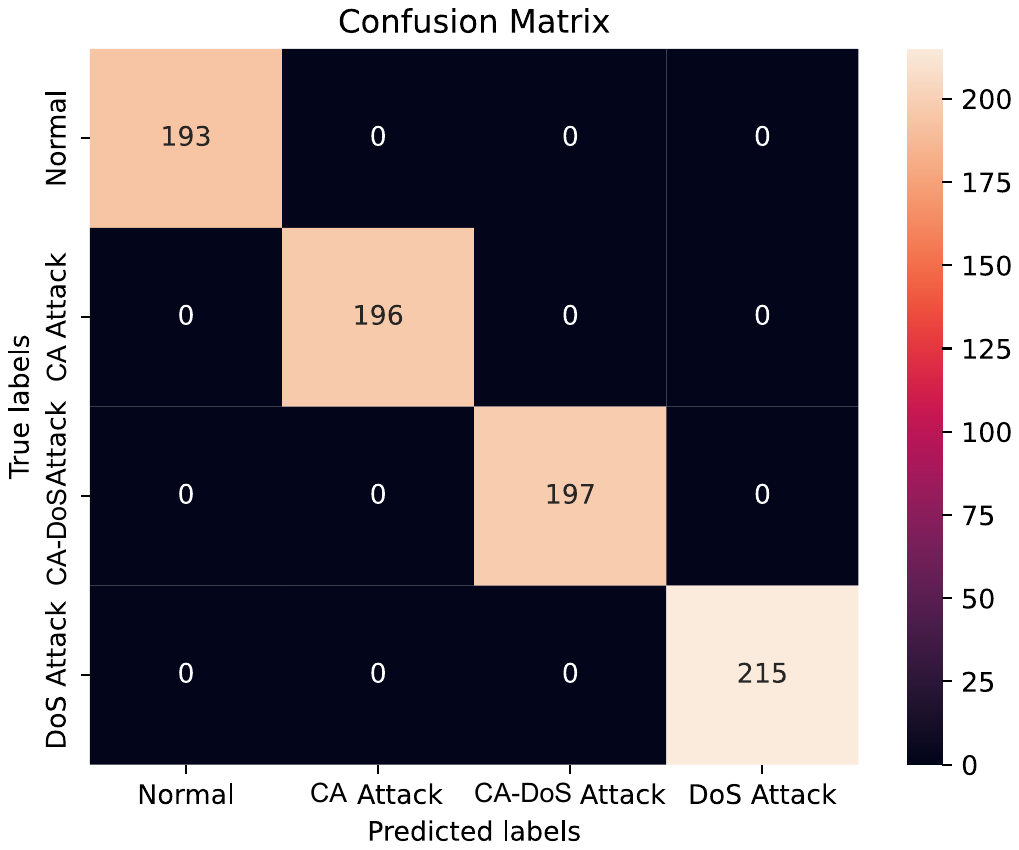}
    \caption{Accuracy= $100\%$.}
    \label{fig:second}
\end{subfigure}
\hfill
\begin{subfigure}{0.47\textwidth}
    \includegraphics[width=\textwidth,trim=0mm 50mm 20mm 50mm, clip=true]{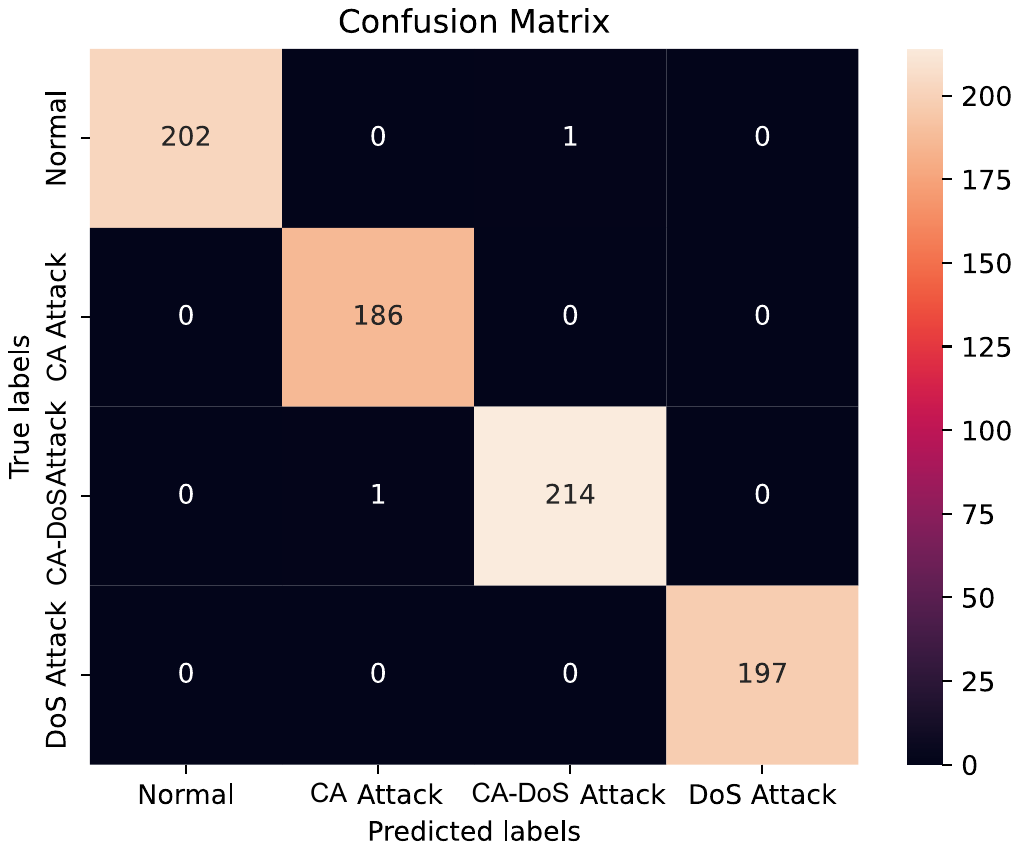}
    \caption{Accuracy= $99.80\%$.}
    \label{fig:third}
\end{subfigure}
\hfill
\begin{subfigure}{0.47\textwidth}
    \includegraphics[width=\textwidth,trim=0mm 50mm 20mm 50mm, clip=true]{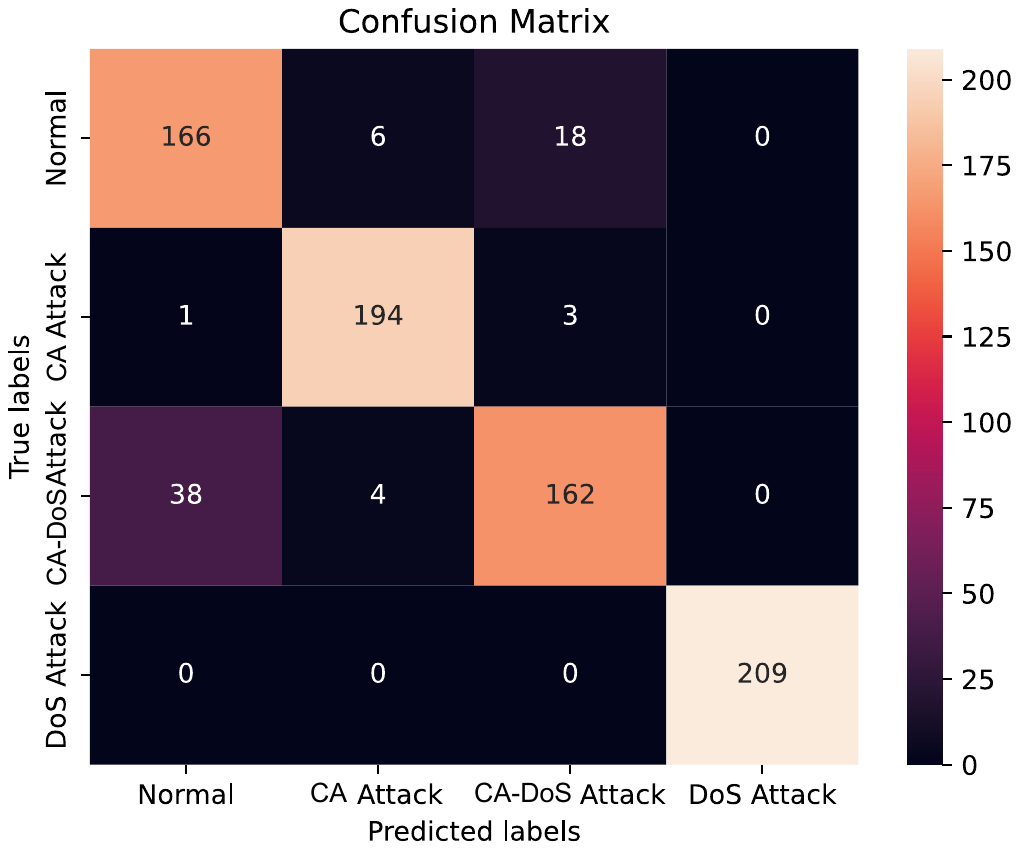}
    \caption{Accuracy= $91.26\%$.}
    \label{fig:third}
\end{subfigure}
\hfill
\caption{The confusion matrix for classifying the types of channel tampering attacks. In (a) $D=40~\text{ km}$ where Eve performs a CA attack at $D_{\text{Eve}}=10~\text{ km}$ and the $\sigma_{\text{RIN,LO}}=0.01$. The attack parameters for the CA attack are $g=1.12$ and $p=1$, for the CA-DoS attack $g=1.12$ and $p=0.94$, and for the DoS attack $g=0.9$ and $p=0.9$. In (b), the same parameters as subfigure (a) but $\sigma_{\text{RIN,LO}}=0.1$. In (c) $D=40~\text{ km}$ where  Eve performs a CA attack at $D_{\text{Eve}}=1~\text{ km}$ and the $\sigma_{\text{RIN,LO}}=0.01$. The attack parameters for the CA attack are $g=1.01$ and $p=1$, for the CA-DoS attack $g=1.01$ and $p=0.99$, and for the DoS attack $g=0.9$ and $p=0.9$. Subfigure (d) is the same as (c) but with $\sigma_{\text{RIN,LO}}=0.1$.}
\label{confusionmatrix}
\end{figure*}
\section{Machine Learning Classification Results}\label{sec_ml}
 After the parameter estimation, a ML model for classifying attacks is used to 
identify the three types of attacks outlined in the previous section. 

\begin{figure*}[!htb]
\centering
\begin{subfigure}{1.1\textwidth}
    \includegraphics[width=\textwidth,trim=25mm 0mm 0mm 0mm, clip=true]{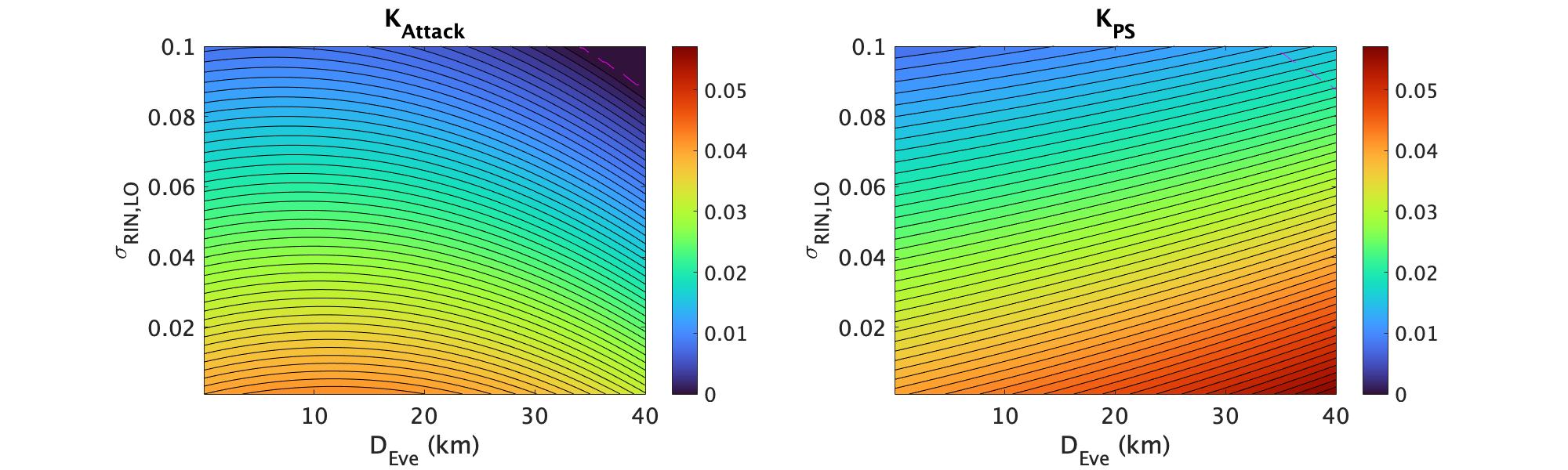}
    \caption{}
    \label{fig:first}
\end{subfigure}
\hfill
\begin{subfigure}{1.1\textwidth}
    \includegraphics[width=\textwidth,trim=25mm 0mm 0mm 0mm, clip=true]{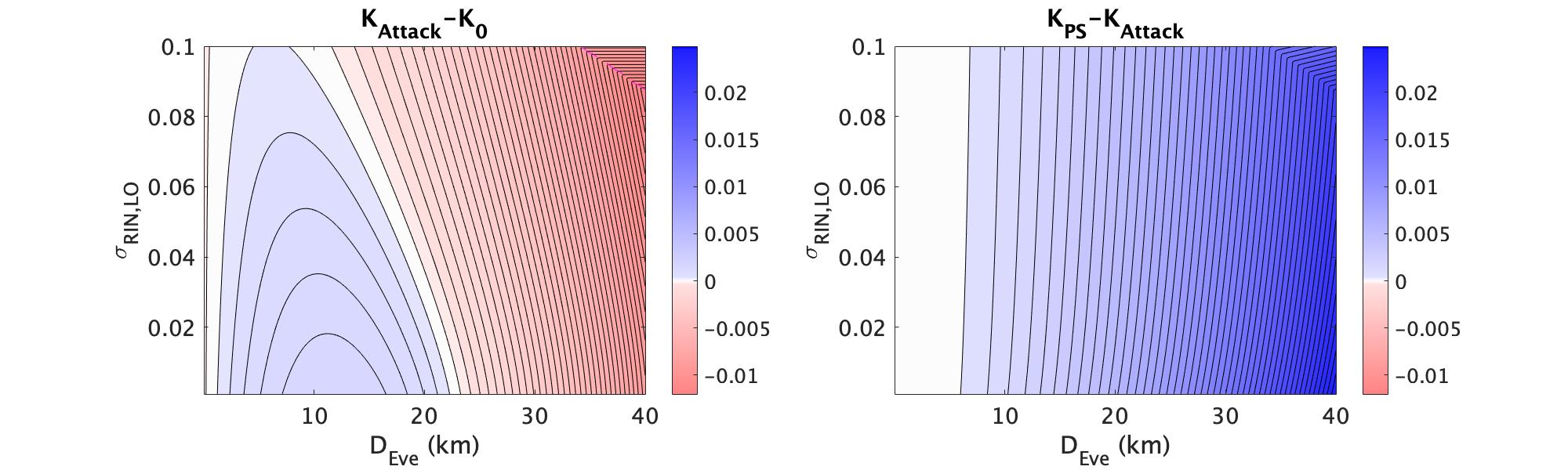}
    \caption{}
    \label{fig:second}
\end{subfigure}       
\caption{In (a) is the SKR (in bits per pulse) under attack (left), and with the mitigation (right) in a $40~\text{km}$ optical fibre. In (b) is the change of the SKR after the CA attack (left) and the SKR improvement (right) plotted against the RIN of the LO and the length of Eve's optical fibre. The block size is $N=10^{12}$, $\varepsilon_{\text{cor}}=\varepsilon_{\text{{PE}}}=\varepsilon_h=\varepsilon_s=10^{-9}$, $p_{\text{ec}}=0.99$ and the number of symbols used for parameter estimation is $m=N/10$. The magenta dashed line is when $K_{\text{Attack}}=0$. $V_A$ is optimized to maximize $K_0$. (Other parameters: $f_{\text{attack}}=0.5$, $\beta=0.9$, $v_{\text{el}}=0.05$, $\eta=0.9$, $\xi_B=0.01$ and $N_0=1$). }
\label{attack1}
\end{figure*}

Our supervised ML model is as follows:
 \begin{itemize}
     \item Based on the distribution $\mathcal{P} (T)$, we generate $N_{Samples} \times M$ samples of $T_i$'s of the normal operating CV-QKD system $T_{\text{norm}}$ in optical fibre and of the three types of attacks $T_{\rom{1},\rom{2},\rom{3}}$ given by the cases in Section \ref{sec_attacks}. The sample size is $N_{Samples}=1000$ and $M=800$. We note that $N_{Samples}$ and $M$ are not the same as the block size or number of symbols used for parameter estimation. These are independent samples from the measurement of the LO intensity.
     
     \item The feature vectors for the four classes $\{ X_{norm}, X_{\rom{1}},X_{\rom{2}},X_{\rom{3}}\}$ where $X$ is the expectation values $\mathbb{E}(T)$ and $\mathbb{E}(\sqrt{T})$ of the sample transmittance values generated i.e. $T_{\text{norm}}$ and $T_{\rom{1},\rom{2},\rom{3}}$. 
     \item The feature vectors are labelled using one-hot encoding $\{\{1,0,0,0\},\{0,1,0,0\},\{0,0,1,0\},\{0,0,0,1\}\}$ for the four classes  $\{ y_{norm}, y_{\rom{1}},y_{\rom{2}},y_{\rom{3}}\}$, respectively.
     \item A ratio of $4:1$ of the shuffled generated data is used for training $\{X_{\text{train}},y_{\text{train}}\}$ and for testing $\{X_{\text{test}},y_{\text{test}}\}$.
     \item We use the \emph{decision tree classifier} model as it is lightweight and requires reduced computational power \cite{decisiontree} to generate predictions $\{X_{\text{predict}},y_{\text{predict}}\}$ on the test data.
 \end{itemize}

In essence, as depicted in Fig. \ref{decisiontree}, the supervised ML model learns from the labelled data of $\mathbb{E}(T)$ and $\mathbb{E}(\sqrt{T})$. It makes a decision based on the features of the data. Based on this knowledge, the type of attack can be predicted from unlabelled data. 

Fig. (\ref{confusionmatrix}) depicts the confusion matrices for classifying the channel tampering attacks using this ML model. We find that when Eve controls the optical fibre starting from Alice for $D_{\text{Eve}}=10~\text{ km}$, the channel tampering attacks are classified with $100 \%$ accuracy for both low noise $\sigma_{\text{RIN,LO}}=0.01$ and high noise $\sigma_{\text{RIN,LO}}=0.1$. However, for $D_{\text{Eve}}=1~\text{km}$, the accuracy drops as the channel parameters are only slightly modified by the attack. It is especially harder for the ML model to distinguish between CA and CA-DoS attacks with a high RIN of LO. For an LO of typical optical bandwidth in optical fibre $\sigma_{\text{RIN,LO}}=0.0014$ \cite{laudenbach}. Hence, the noise of the LO in typical CV-QKD implementations are rarely of the high noise regime.

\section{Post-Selection of Classified Attack}
\label{sec_postselection}
If Alice and Bob continue their protocol without identifying the type of attack and frequency of attack $f_{\text{attack}}$, the new channel parameters, as measured by Alice and Bob, are the weighted average of the attacked and unattacked probability distributions. Therefore,

\begin{equation}
\begin{split}
   \mathbb{E}(\sqrt{T})&=f_\text{attack} p \sqrt{g T_0}+(1-f_\text{attack}) \sqrt{T_0},\quad \text{and}\\
    \mathbb{E}(T)&=f_\text{attack} p g T_0+(1-f_\text{attack}) T_0.
\end{split}
\end{equation}
The channel parameter estimators are given by Eq. (\ref{parameterattack}) and explicitly shown in Appendix \ref{appedixB}. Based on these channel parameters, we denote $K_{\text{Attack}}$ as the `attacked SKR' where Bob carries on with the channel estimation without knowledge of the attack. 

However, with our ML model, Alice and Bob can classify the type of channel tampering attack and tag the quadrature data. The post-selected SKR is binned into the unattacked sub-channel and attacked sub-channel: 

\begin{equation}
    K_{\text{PS}}=(1-f_{\text{attack}}) K_0(T_0,\xi_0)+f_{\text{attack}} K(T',\xi'),
    \label{kps}
\end{equation} where $K_0(T_0,\xi_0)$ is the unattacked SKR and $K(T',\xi')$ is the attacked SKR where $T'$ and $\xi'$ are the estimators given by Eq. (\ref{CTattack}). The above formula applies to both the asymptotic limit and finite-size effects. In this section, we will see how the attack affects the SKR, and also assess the performance of the mitigation strategy.

The chosen parameters for our CV-QKD system are $\beta=0.9$ for the reconciliation efficiency, the baseline excess noise $\xi_B=0.01$, the electronic noise $v_{el}=0.05$, the detector efficiency $\eta=0.9$ and the shot noise $N_0=1$. These parameters are based on experimental values from \cite{huangsci}. The block size is $N=10^{12}$, $\varepsilon_{\text{cor}}=\varepsilon_{\text{{PE}}}=\varepsilon_h=\varepsilon_s=10^{-9}$, $p_{\text{ec}}=0.99$ and the number of symbols used for parameter estimation is $m=N/10$. Alice's modulation variance $V_A$ is optimized w.r.t. the unattacked channel parameters and falls in the range $V_A=\{2.79,2.89\}$. 
\begin{figure}[t!]
\centering
    \includegraphics[width=0.33\textwidth]{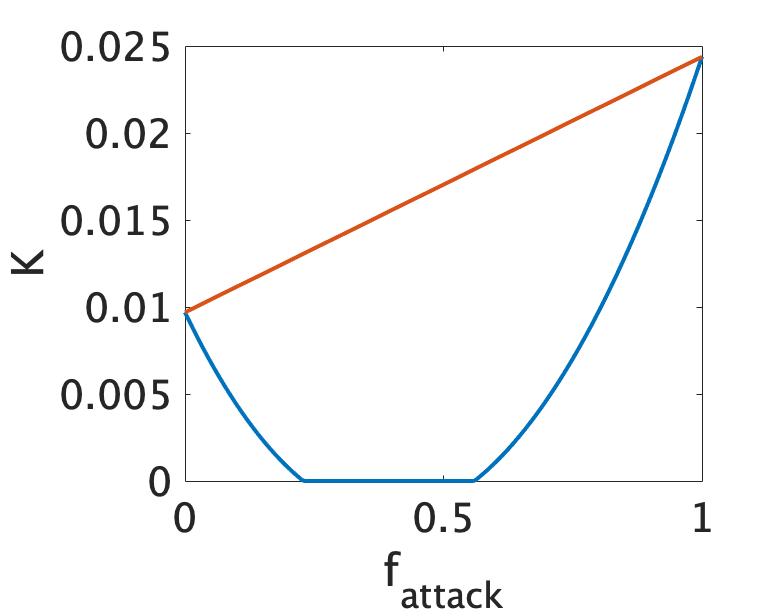}
\caption{SKR as a function of $f_\text{attack}$ for $K_\text{Attack}$ (blue) and $K_{\text{PS}}$ (red) with $D_{\text{Eve}}=39~\text{ km}$ and $\sigma_{\text{RIN,LO}}=0.098$. }
\label{frequency}
\end{figure}

\begin{figure*}[t!]
\centering
\begin{subfigure}{1.1\textwidth}
    \includegraphics[width=\textwidth,trim=25mm 0mm 0mm 0mm, clip=true]{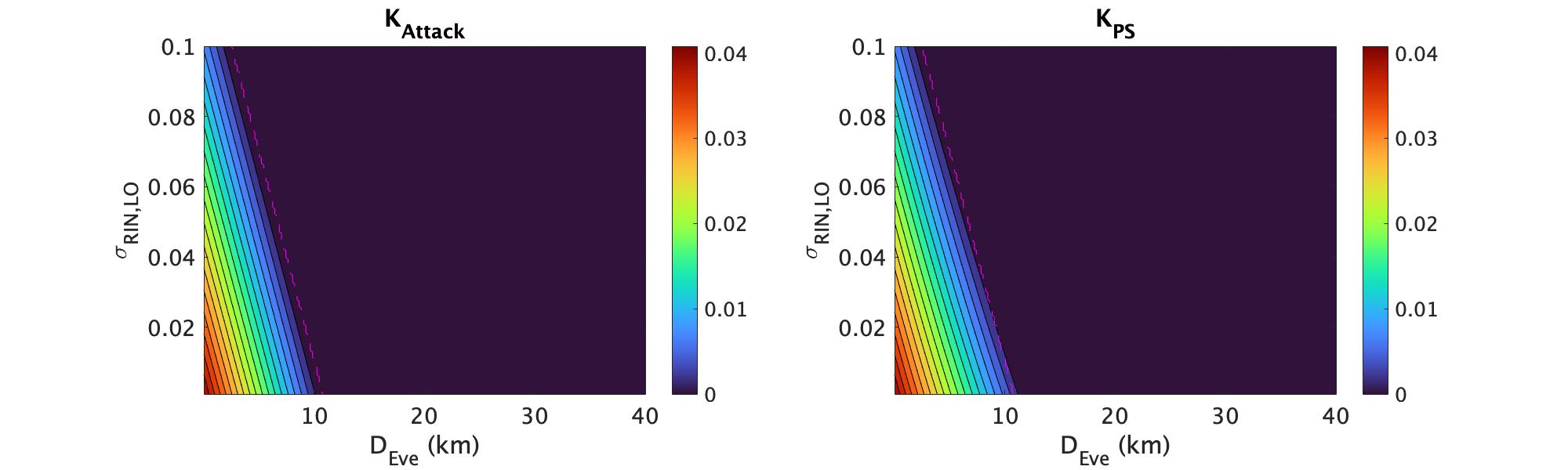}
    \caption{}
    \label{fig:second}
\end{subfigure}  
\hfill
\begin{subfigure}{1.1\textwidth}
    \includegraphics[width=\textwidth,trim=25mm 0mm 0mm 0mm, clip=true]{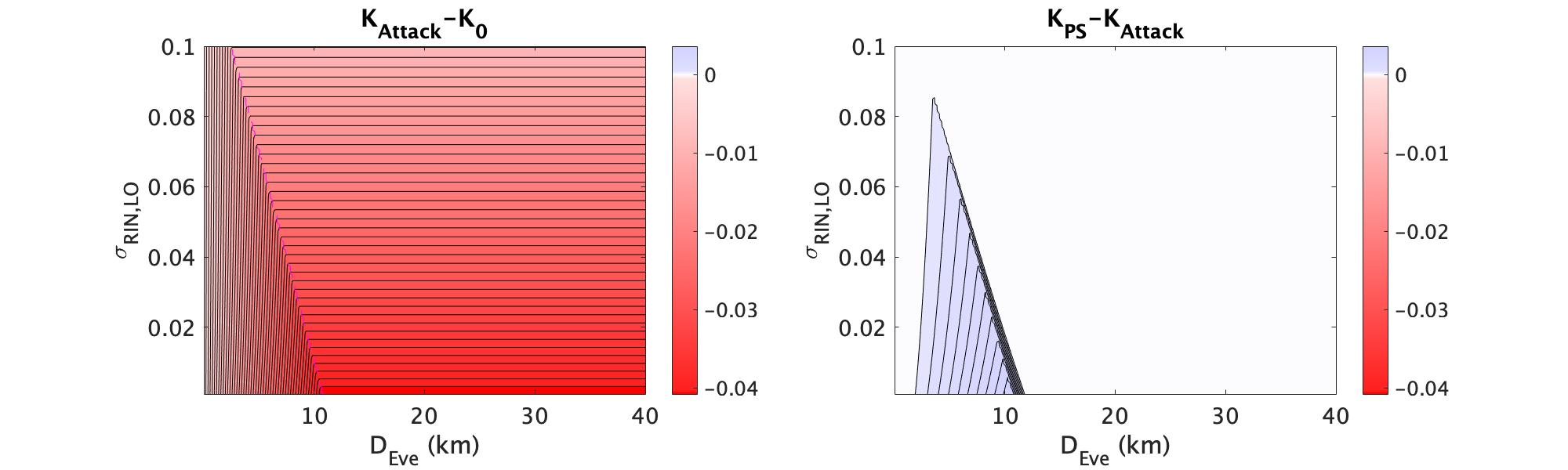}
    \caption{}
    \label{fig:second}
\end{subfigure} 
\caption{
CA-DoS attacks with parameters the same as Fig. \ref{attack1}. }
\label{attack2}
\end{figure*}
In Fig. \ref{attack1} (a), the SKR is plotted for $\sigma_{\text{RIN,LO}}$ and $D_{\text{Eve}}$. It can be seen that the SKR with the CA attack $K_{\text{Attack}}$ in the left subfigure has a region in the top right where there is zero SKR. As seen in the right subfigure, the SKR after the mitigation $K_{\text{PS}}$ then solves this issue by achieving a positive SKR in this region and increasing the SKR overall. In Fig. \ref{attack1} (b), we can see the effect of the CA attack on the SKR. Compared to the unattacked SKR $K_0$, the attacked SKR $K_\text{attack}$ increases in the blue region, while decreases in the red region. After mitigation, as seen in the right subfigure, the SKR is better for all parameters. The SKR improvement is up to $0.024 \text{ bits/pulse}$. Of particular significance is the region at which the mitigation increases from zero to non-zero SKR. In Fig. \ref{frequency}, we plot the SKR improvement as a function of Eve's frequency of attack for a point in this region. It can be seen that the SKR always drops and is zero for $f_{\text{attack}}=0.25 \text{ to  } 0.6$.


In the left-hand side of Fig. \ref{attack2} it can be seen that the CA-DoS attack significantly reduces the SKR. On the right-hand side of the figure, our mitigation strategy is found to improve the SKRs under attack with Eve between $2-12 \text{ km}$ in the triangular region shown in the right subfigure of  Fig. \ref{attack2} (b). The SKR advantage is up to $0.002 \text{ bits/pulse}$. Despite this marginal improvement, this attack is more detrimental to communication compared to the CA attack. 

For CA attacks with a bandwidth or pulse rate of $100~\text{ MHz}$, the SKR in bits/s improves on average by $6~\text{ kb/s}$ which is comparable to the average SKR of $20~\text{ kb/s}$. For CA-DoS attacks, the improvement in the region of non-zero SKR is on average $1.6~\text{ kb/s}$ compared to the average SKR of $15~\text{ kb/s}$. For state-of-the-art repetition rates of $1\text{ GHz}$, these quotes SKR values would increase an order of magnitude. 

\emph{False classification}. We repeat here that the ML classification is for binning of the data to optimize the SKR, and the quadrature data contains all information to bound Eve’s information as described in Section \ref{sec_quantum_security}. The channel tampering attack modifies the transmittance-- a parameter assumed to be controlled by Eve in the security analysis-- to disrupt communication. This contrasts with Eve exploiting the hardware-specific implementation to facilitate a side-channel attack. Hence, false classifications would simply change the statistics of $T$ and $\xi$ but not the security of the coherent state protocol. 

To further this point, we explore the scenario where there would be a CA attack (a positive) but the ML model falsely classifies this attack as a normal channel without Eve tampering the channel i.e. a false negative (FN). For example, in Fig. \ref{confusionmatrix} d) this happens $1$ out of $800$ times in this simulation. Upon this false classification, the block of correlated $\braket{X_A  X_B}$ is binned into the normal data wrongly with $T_{\text{Wrong}}=g T_0$ about $1$ out of $800$ times for the total block size $N=10^{12}$. 

The penalty for this false classification\footnote{ We note that there is also the possibility of falsely classifying under another attack such as the CA-DoS or DoS attack, leading to less efficient parameter estimation.} is that the estimate for the transmittance from the binned normal data is  $\hat{T}=(E(\sqrt{T}))^2=(799 \sqrt{T_0}  + \sqrt{g T_0})^2/800^2=(T_0 (799+\sqrt{g})^2)/800^2$ which is slightly higher than $T_0$. Whereas if the normal data is falsely classified as a CA attack $3$ in $400$ times i.e. false positives (FP), the estimated transmittance is $\hat{T}=(E(\sqrt{T}))^2=(397 \sqrt{g T_0}   + 3 \sqrt{T_0})^2/400^2=(g T_0 (397+3/\sqrt{g})^2)/400^2$ which is slightly less than $g T_0$. Hence false classifications do not change the security of the protocol but rather reduce the efficiency of the parameter estimation. 

In this example, despite some of the quadrature data tagged incorrectly, the security is dependent on the transmittance of the unattacked and attacked sub-channels with true negative + false negative and true positive + false positive binned data, respectively. Although the SKR may not be as high as the most accurate ML model, the security of the coherent state protocol remains unchanged. Hence, using Eq. \ref{kps}, we can continue to employ the post-selection method. 


\emph{Fast ML classification}. The number of samples in our ML model is $N_\text{Samples}=1000$ and $M=800$. The mean time for the ML model using the scikit-learn Python libraries on a 12th Gen Intel(R) Core(TM) i7-1270P 2.20 GHz with 32.0 GB RAM to make a prediction is $0.0042~\text{s}=4.2 \text{ ms}$. For a repetition rate of $100~\text{MHz}$, this means every batch of samples takes only $0.1~\text{ ms}$. Therefore, there is a slight bottleneck for processing and tagging Bob's quadrature data in high repetition systems. Dedicated options including the use of Field-Programmable Gate Arrays (FPGAs) and Graphics Processing Units (GPUs) could optimize this process. Hence, this necessitates lightweight and fast ML models for detection of channel tampering attacks in real-time.




\section{Discussion}\label{sec_discussion}
Our results indicate that the classification of channel tampering attacks 
enhances the SKR under attack of CV-QKD. 
We have devised an attack model, where Eve can perform a channel amplification attack.
In the attack, Eve can amplify the transmittance of the channel and counterintuitively reduce the SKR as seen in Fig. \ref{attack1}. The channel amplification attack can be understood as Eve disturbing the channel to deviate the system from the optimal performance. At $D_{\text{Eve}}=40~\text{ km}$, the amplification of the CA attack is $g=1.58$ and hence the average $\hat{T}=\frac{(\sqrt{1.58}+1)^2 T_0}{4} \approx 1.27 T_0$. After the mitigation, the subchannels are $T_{\text{unattacked}}=T_0$ and $T_\text{attacked}=1.58 T_0 $. Since Alice optimized her modulation variance $V_A$ for $T_0$ and $\xi_0$, the SKR is the maximum for the unattacked subchannel $(T_0,\xi_0)$ multiplied by the factor $1-f_\text{attack}$. Hence the improvement of the SKR is especially seen when the attack nullifies the SKR in Fig. \ref{attack1} a) at $40~\text{ km}$ and at high RIN of the LO noise can be explained mainly due to this SKR contribution. Conversely, when Eve hijacks a short-length fibre as seen in the left subfigure of Fig. \ref{attack1} b) (up to $23~\text{ km}$), the CA attack slightly improves the SKR because of the small amplification of the channel. As seen in the right-hand figure, the mitigation does not improve the SKR below approximately $6~\text{ km}$ because the $V_A$ for the attacked sub-channel remains close to the optimal $V_A$ of the unattacked sub-channel. In addition, when post-selecting into attacked and unattacked, the blocks decrease to $f_\text{attack} N$ and $(1-f_\text{attack}) N$ and hence the SKR decreases due to finite-size effects. However, for the chosen block-size $N=10^{12}$ there is no discernible detriment to the SKR. Nonetheless, post-selection can reduce the SKR for small block-sizes as seen in Appendix \ref{smallblock}.

As demonstrated by our numerical simulations in Fig. \ref{confusionmatrix}, the accuracy of our ML classifier is $100 \%$ in operational environments with typical noise. This accuracy drops to $91.4 \%$ in high noise environments. 
Our numerical calculations in Figs \ref{attack1} and \ref{attack2} show that there is insignificant SKR improvement 
of correct CA or CA-DoS attack classification when Eve is near to Alice i.e. for $D_{\text{Eve}}\le 1~\text{ km}$. 
Coincidentally this is when the ML model is least accurate as Eve barely tampers with the channel. Hence, a false positive or false negative classification would likewise not be detrimental. Nonetheless, future work could consider spectral estimation techniques for detecting low rate of DoS attacks in this parameter regime as studied in \cite{ldos}. 


As confirmed numerically in Fig. \ref{attack1}, we find that our mitigation strategy works especially well for some parameter regimes. We can explain its effectiveness by the following statement: the information-theoretic security of the Gaussian modulated CV-QKD relies on Gaussian optimality, which explains the overestimation of Eve's information due to the averaging of channel parameters if Alice and Bob do not post-select their data. 
Averaging over the channel parameters loses information advantage over Eve (i.e. SKR), but it still remains secure as Eve's information is overestimated (the average of two Gaussians broadens the Gaussian distribution). Post-selection is effective because the parameter estimation of $T$ and $\xi$ is binned into sub-channels, rather than averaging over the channel parameters. This post-selection technique, also known as data clusterization, has been applied to free-space CV-QKD in \cite{Ruppert_2019, deq}, and here we find an application to optical fibre. Nonetheless, the type of post-selection in our method involves only the binning of the data, whereas other methods of post-selection that emulate noiseless amplifiers were shown to improve SKR \cite{postselection2020}. It is left to future work to determine whether this type of post-selection on the attacked binned data can further help mitigate the channel tampering attacks.

In practice, the CA attack that we propose here involves the coupling and decoupling of the signal and LO from a standard optical fibre to an ultra-low loss fibre. In our results, we assumed no additional optical loss. Current technology does not allow this as there is an inevitable loss through the optical switches and coupling between fibres \cite{Zhang_2019}. Nonetheless, even for a less-than-perfect CA attack, a mitigation strategy based on ML and post-selection, as we have shown here, would help improve the SKRs under attack. 
Although we have focused here only on the specifications of a metropolitan scale system in optical fibre, the CA attack could be adapted to free-space CV-QKD and longer distances. A CA attack in free-space CV-QKD may take the form of a focusing lens to reduce the broadening of the beam, thus modifying $T$.  

One of the limitations of our ML classification model is that the model only learns from a specific set of possible channel tampering attacks. We have not yet tested on an unknown attack. Also, our model is not universal
in detecting side-channel attacks as done in  \cite{universalattack}. Hence, there is a tremendous need to include all known attacks including the channel tampering attacks in the next generation of ML models for attack detection. 

Furthermore, our system is set up to monitor the intensity fluctuations of LO. 
Our assumption is that LO and signal are affected by the eavesdropper equally. 
However, as the LO and signal-to-noise ratio can be modified independently, 
Eve can perform an intercept-and-resend attack \cite{joug2013, LOintensityDM}. 
Other attacks, such as the wavelength attack, share similar characteristics to the CA attack.
The similarity is from the fact that the transmittance of an optical component is modified \cite{huang2013, huang2014}.
The difference is that the CA attack modifies the channel rather than the shot noise $N_0$. {Finally, we remark on the channel tampering attack in CV-QKD systems with locally generated LOs at Bob. In these systems, the reference pulse, which is usually a strong pulse is used to synchronize the quadrature reference of Alice and Bob \cite{marie,Wang:18}. In our analysis of the channel tampering attack, we implicitly assumed that there are no other modifications of the reference pulse or LO. However, Eve can modify the reference pulse to exploit security loopholes \cite{shaoyun}, compared to the channel tampering attack on the signal pulse. This leads to the following discussion on the need for universal countermeasures against side-channel attacks in addition to physical attacks.}

A universal ML detection scheme would include both quadrature data from the signal and intensity 
measurements of LO to monitor all the aforementioned attacks that independently affect 
the signal and LO. Despite significant advancements in CV-QKD and its commercial availability, vulnerabilities persist due to imperfections in hardware implementation. This leads to various side-channel attacks such as local oscillator (LO) intensity fluctuation, wavelength, calibration, and saturation attacks \cite{sidechannel, joug2013,xma, huang2013, huang2014, xma2013, qin2016, Zhao_2019, shao2022}. Consequently, defense strategies necessitate meticulous monitoring of transmission parameters, albeit adding complexity to the CV-QKD design \cite{xma2014manipulate}. We leave it to future research to develop a universal defense strategy 
that includes all types of attacks including the novel threat that we have devised here.

We note that amplifying the signal can expose CV-QKD implementations to side-channel attacks as recently considered in Ref. \cite{CVQKDopticalamplifier}.  Authors motivate an attack based on modifying the magnetic field of an eridium-doped fibre amplifier (EDFA). The effect of this vulnerability is that Eve can perform a partial intercept-and-resend attack without being detected. Despite also amplifying the signal, the difference is that in the channel tampering attack, it is assumed that Eve does not exploit this side channel to perform an additional attack. Nonetheless, it would be interesting as a future work to consider the feasibility of intercept-resend attacks hidden behind channel tampering attacks. We also note the importance of optimal quantum attacks in practical optical fibre setups recently considered in Ref. \cite{liu2023security}. Authors in Ref. \cite{liu2023security} explore different types of optical fibre between Alice and Bob as well as part of Eve's teleportation-based collective attack.  In our paper, we do not consider Eve’s optimal quantum attack strategies in practical settings and we assume the much more “paranoid” security approach where the eavesdropper employs the most powerful quantum attack with infinite resources and no loss (this is the red curve in Fig. 4 of \cite{liu2023security}). In some limits, if Eve’s EPR source suffers no loss then this will lead to the CA attack if the optical fibre between Alice and Bob is replaced with fibre of decreased loss. Contrary to the motivations of this paper, the channel tampering attack in our paper is proactive and is aimed to subtly disrupt communications between Alice and Bob. 


\section{Conclusion}\label{sec_conclusion}
In this work, we devised a new physical attack involving the amplification of the channel in an implementation of CV-QKD in optical fibre. The effect of the CA attack is not trivial, requiring a detection and mitigation strategy. Furthermore, we developed an ML model that learns the features of the LO intensity and classifies the type of attack on the channel.  The knowledge of the existence of diverse channel tampering attacks improves the efficiency of CV-QKD. Subsequently, we set up a mitigation strategy that makes use of ML classification to post-select quantum data and improve the SKRs under attack. Our results are important for ensuring the robustness of CV-QKD systems.


\bibliography{sample}

\section*{Acknowledgements}
This research has been supported by the Commonwealth Scientific and Industrial Research Organization (CSIRO) and by the Office of the Chief Scientist of CSIRO Impossible Without You Program. We also thank the reviewers for their valuable comments to improve this article. 

\section*{Author contributions statement}
S.K. wrote the manuscript, performed the calculations, devised the idea, and generated all results. C.T., M.S., H.S., J.P., and S.C. all helped proofread and edit the manuscript. 

\appendix
\section*{Appendices}
\addcontentsline{toc}{section}{Appendices}
\renewcommand{\thesubsection}{\Alph{subsection}}
\subsection{Parameters under attack}
\label{modified}
The estimate for the square root of the transmittance is straightforward to calculate from the equations $\hat{t}=\frac{\frac{1}{m}\sum^{m-1}_{i=0} x_{A,i} x_{B,i}}{\frac{1}{m}\sum^{m-1}_{i=0} x^2_{A,i}}$ and the linear model of $x_B=\hat{t} x_A+z_R$ where $\hat{t}=\sqrt{\eta} \sqrt{\hat{T}}$ with $\hat{T}$ no longer constant. It then follows that $\frac{1}{m}\sum^{m-1}_{i=0} x_{A,i} x_{B,i}=\sqrt{\eta} \mathbb{E}(\sqrt{T}) V_A$ where it was used that $ \frac{1}{m}\sum^{m-1}_{i=0} x_{A,i} z_r=0$. i.e. the noise is centered at $0$ and that $V_A=\frac{1}{m}\sum^{m-1}_{i=0} x^2_{A,i}$. Hence $\hat{t}=\mathbb{E}(\sqrt{T})$.
For the excess noise, we note the Gaussian noise of the channel is:
\begin{equation}
\begin{split}
    \hat{\sigma}^2_R&=\frac{1}{m} \sum^n_{i=0} (x_{B,i}-\hat{t} x_{A,i})^2\\
    &=\frac{1}{m} \sum^n_{i=0} (x^2_{B,i}-2 \hat{t} x_{A,i} x_{B,i} +\hat{t}^2 x_{A,i}^2) \\
    &=\mathbb{E}(X_B^2)-2 \hat{t} \mathbb{E}(X_B X_A)+\hat{t}^2 \mathbb{E} (X_A^2) \\
    &=\eta \mathbb{E}(T) (V_A+\xi)+N_0+v_{\text{el}}- \hat{t}^2 V_A,
\end{split}
\end{equation} where we have used $\mathbb{E}(X_B^2)=\eta \mathbb{E}(T) (V_A+\xi)+N_0+v_{\text{el}}$ and $\mathbb{E}(X_B X_A)=\mathbb{E} (\sqrt{\eta T}) V_A=\hat{t} V_A$. Since $\sigma^2_R=\eta T \xi+v_{\text{el}}+N_0$ then the estimated noise $\hat{\sigma}^2_R=\eta \hat{T} \hat{\xi}+v_{\text{el}}+N_0$ where rearranging for $\hat{\xi}$:
\begin{equation}
\begin{split}
    \hat{\xi}&=\frac{\hat{\sigma}^2_R-v_{\text{el}}-N_0}{\eta \hat{T}}\\
    &=\frac{\mathbb{E}(T)}{\hat{T}} (V_A+\xi)- V_A,
\end{split}
\end{equation} where we have used that $\eta \hat{T}=\hat{t}^2$.

\subsection{Coherent-state protocol with heterodyne detection}
\label{appendixgg02}
 In a Gaussian thermal-loss channel, the quadrature covariance matrix between Alice and Bob is \cite{raulsanchez}:
\begin{equation}
\begin{split}
&\gamma_{\rm AB}=
\begin{pmatrix}
a\mathbb{I} & c\sigma_{z} \\ c\sigma_{z} & b\mathbb{I}
\end{pmatrix}
=
\begin{pmatrix}
\gamma_\text{A} & \sigma_{\text{AB}} \\ \sigma_{\text{AB}} & \gamma_\text{B}
\end{pmatrix} = \\
&
\begin{pmatrix}
(V_\text{A}+1)\mathbb{I} & \sqrt{\eta(V_\text{A}^2+2V_\text{A})}\sigma_{z} \\ \sqrt{\eta(V_\text{A}^2+2V_\text{A})}\sigma_{z} & V_B\mathbb{I}
\end{pmatrix},
\end{split}
\label{covariance}
\end{equation}
where $V_\text{A}$ and $V_B$ are the variances measured by Alice and Bob (respectively), $\mathbb{I}=\text{diag}(1,1)$ is the unity matrix and $\sigma_{z}~=~\text{diag}(1,-1)$ is the Pauli-Z matrix. 
We choose heterodyne detection (also known as ``no switching") at Bob, in which Bob measures both $X$ and $P$ quadrature measurements. 


For heterodyne detection by Bob, the mutual information $I_{AB}$ in a thermal-loss channel is\cite{raulsanchez}
\begin{equation}
\begin{split}
    I_{\text{AB}}&=\log_2{\Bigg( \frac{V_\text{B}+1}{V_{\text{B}|\text{A}^M}+1} \Bigg)}
\end{split}
\label{mutual}
\end{equation} where $V_\text{B}$ is Bob's variance and $V_{\text{B}|\text{A}^M}=b-c^2/(a+1)$ is Bob's variance conditioned on Alice's heterodyne measurement. 

The Holevo information between Bob and Eve for the collective attack is given by 
\begin{equation}
\chi_{\text{EB}}=S(\text{E})-S(\text{E|B}),
\label{eve}
\end{equation} where $S(\text{E})$ is Eve's information and $S(\text{E|B})$ is Eve's information conditioned on Bob's measurement. In Eve's collective attack, Eve holds a purification of the state between Alice and Bob with entropy given by
\begin{equation}
S(\text{E})=S(\text{AB})=G[(\lambda_1-1)/2]+G[(\lambda_2-1)/2],
\label{pure}
\end{equation} where $G(x)=(x+1) \log_2{(x+1)}-x \log_2{x}$ and $\lambda_{1,2}$ are the symplectic eigenvalues of the covariance matrix $\gamma_{\text{AB}}$ given by
$\lambda_{1,2}^2=\frac{1}{2} [\Delta \pm \sqrt{\Delta^2-4\mathcal{D}^2}]$, where 
$\Delta=\text{Det}(\gamma_\text{A})+\text{Det}(\gamma_\text{B})+2\text{Det}(\sigma_{\text{AB}})$, and
$\mathcal{D}=\text{Det}(\gamma_{\text{AB}})$.

$S(\text{E|B})=S(\text{A}|x_B,p_D)$ is the information obtained by Eve conditioned on Bob's heterodyne measurement result $x_\text{B}$ and the auxiliary mode $p_\text{D}$ \cite{raulsanchez}. 

The covariance matrix of Alice after a projective measurement by Bob's heterodyne detection is
\begin{equation}
\gamma_\text{A}^{\text{out}}=\gamma_\text{A}-\sigma_{\text{AB}}(\gamma_\text{B}+\mathbb{I})^{-1} \sigma_{\text{AB}}^T,
\end{equation} where $\sigma_{\text{AB}}=c \sigma_Z$. The conditional Von Neumann entropy is
\begin{equation}
S(\text{A}|x_\text{B},p_\text{D})=G[(\lambda_3-1)/2],
\label{cond}
\end{equation} where the symplectic eigenvalue $\lambda_3$ is
\begin{equation}
\lambda_3=a-c^2/(b+1).
\label{lambda}
\end{equation} 
\subsection{Weighted average of the channel parameters}
\label{appedixB}
Eve subsequently performs the channel tampering attack for a portion of the keys $f_{\text{attack}}$. For the transmittance, the expectations values are:

\begin{equation}
    \begin{split}
        \mathbb{E}(\sqrt{T})&=\int{\sqrt{T} (f_{\text{attack}} \mathcal{P}_{\text{attack}}}(T)+(1-f_\text{attack})\mathcal{P}(T))dT, \\
        &=f_\text{attack} p \sqrt{g T_0}+(1-f_\text{attack}) \sqrt{T_0} \\
         \mathbb{E}(T)&=\int{T (f_{\text{attack}} \mathcal{P}_{\text{attack}}}(T)+(1-f_\text{attack})\mathcal{P}(T))dT. \\
        &=f_\text{attack} p g T_0+(1-f_\text{attack}) T_0
    \end{split}
\end{equation}
Therefore, the new channel parameters are:
\begin{equation}
    \begin{split}
        \hat{T}&=(\mathbb{E}(\sqrt{T}))^2\\
        &=T_0(f_\text{attack}^2 p^2 g +2 (1-f_\text{attack}) f_\text{attack} p \sqrt{g} +(1-f_\text{attack})^2), \\
        \hat{\xi}&=\frac{\mathbb{E}(T)}{\hat{T}} (V_A+\xi)-V_A \\
        &=\frac{f_\text{attack} p g +(1-f_\text{attack}) }{(f_\text{attack}^2 p^2 g +2 (1-f_\text{attack}) f_\text{attack} p \sqrt{g} +(1-f_\text{attack})^2)} \\
        &\times (V_A+\xi)-V_A.
    \end{split}
\end{equation}

\begin{widetext}
\subsection{Mitigation for small block-size}
In this Appendix, we show the finite-size limitation of the simple post-selection method. For small block-sizes, as can be seen in Fig. \ref{smallblockfigure}, the mitigation strategy is effective for $D_{\text{Eve}}$ above $20$ km and below $\sigma_{RIN,LO}=0.05$ but is less effective below $20~\text{ km}$. The reason for this last observation is that Bob is reducing the block-size for the sub-channels. In this case, by $f_{\text{attack}}=1/2$. A potential solution to this is to determine the trade-off between this mitigation strategy and transmittance averaging by weighing the data binning of the sub-channels. 
\label{smallblock}
\begin{figure*}[t!]
\centering
\begin{subfigure}{1\textwidth}
    \includegraphics[width=\textwidth,trim=0mm 0mm 0mm 0mm, clip=true]{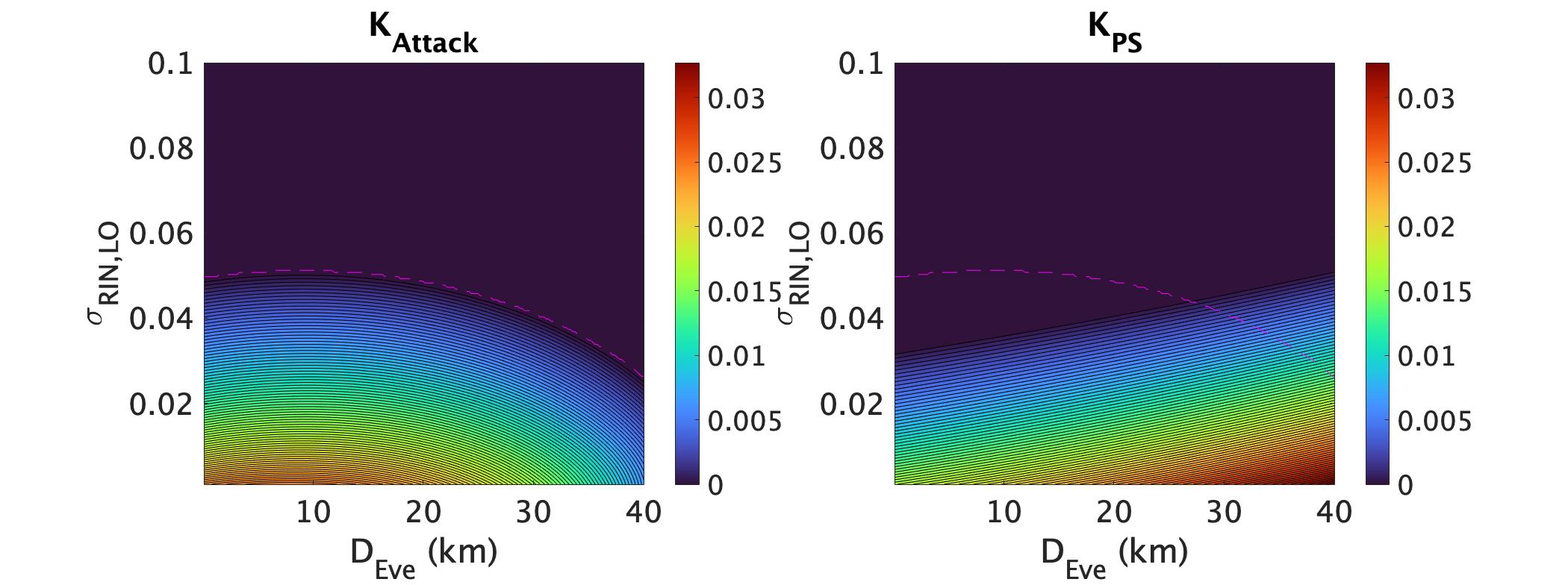}
    \caption{}
    \label{fig:second}
\end{subfigure}  
\hfill
\begin{subfigure}{1\textwidth}
    \includegraphics[width=\textwidth,trim=0mm 0mm 0mm 0mm, clip=true]{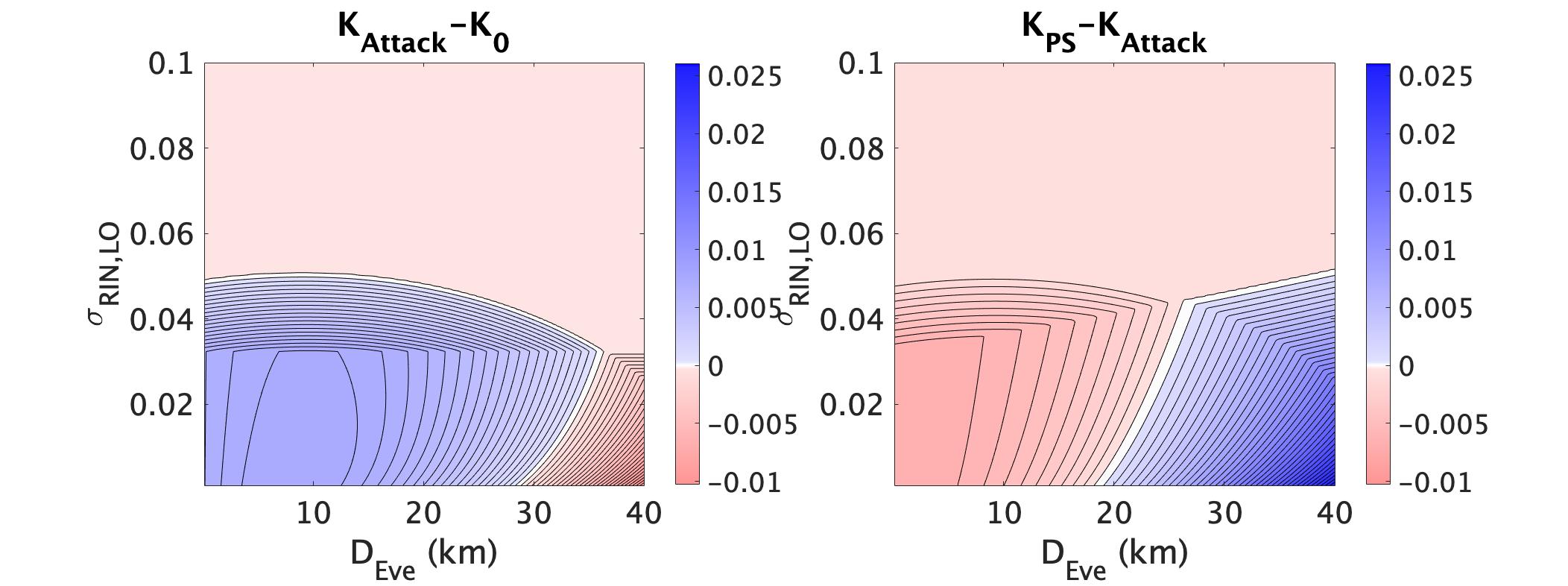}
    \caption{}
    \label{fig:second}
\end{subfigure} 
\caption{
CA attacks with $N=10^8$ block size and all other parameters the same as Fig. \ref{attack1}. }
\label{smallblockfigure}
\end{figure*}
\end{widetext}
\end{document}